\documentclass[prx,nofootinbib,superscriptaddress,showpacs,twocolumn,longbibliography]{revtex4-2}
\usepackage{graphicx}
\usepackage{amsmath,amssymb,physics}
\usepackage{bbold} 
\usepackage{xcolor}
\usepackage{ulem}
\begin{document}
\title{Intertwining Markov Processes via Matrix Product Operators} 
\begingroup
\renewcommand\thefootnote{}
\footnotetext{\emph{\textbf{Notes.}}—The authors of this paper were ordered alphabetically.}
\endgroup
\author{Rouven Frassek}
\email{rfrassek@unimore.it}
\affiliation{University of Modena and Reggio Emilia,
Department of Physics, Informatics and Mathematics,
Via G. Campi 213/b, 41125 Modena and INFN, Sezione di Bologna, Via Irnerio 46,
40126 Bologna, Italy}

\author{Jan de Gier}
\email{jdgier@unimelb.edu.au}
\affiliation{School of Mathematics and Statistics, The University of
Melbourne, VIC 3010, Australia}

\author{Jimin Li}
\email{Corresponding Author: jl939physics@gmail.com}
\affiliation{Centre for Quantum Information and Foundations, Department of Applied Mathematics and Theoretical Physics,
University of Cambridge, Wilberforce Road, Cambridge CB3 0WA, United Kingdom}

\author{Frank Verstraete}
\email{fv285@cam.ac.uk}
\affiliation{Centre for Quantum Information and Foundations, Department of Applied Mathematics and Theoretical Physics, University of Cambridge, Wilberforce Road, Cambridge CB3 0WA, United Kingdom}
\affiliation{Department of Physics and Astronomy, Ghent University, Krijgslaan 281, S9, B-9000 Ghent, Belgium}
\begin{abstract}
{Duality transformations reveal unexpected equivalences between seemingly distinct models. We introduce an out-of-equilibrium generalisation of matrix product operators to implement duality transformations in one-dimensional boundary-driven Markov processes on lattices. In contrast to local dualities associated with generalised symmetries, here the duality operator intertwines two Markov processes via generalised exchange relations and realises the out-of-equilibrium duality globally. We construct these operators exactly for the symmetric simple exclusion process with distinct out-of-equilibrium boundaries. In this case, out-of-equilibrium boundaries are dual to equilibrium boundaries satisfying Liggett's condition, implying that the Gibbs–Boltzmann measure captures out-of-equilibrium physics when leveraging the duality operator. We illustrate this principle through physical applications.
} 
\end{abstract}
\maketitle

\emph{\textbf{Introduction.}}—Dualities have long been used to define equivalences between models and have found wide applications in the theory of strongly correlated many-body systems in equilibrium, ranging from the identification of critical points in lattice models of statistical mechanics to topological phases of interacting quantum spin chains~\cite{kramers1941statistics, lootens2025entanglement,PhysRevB.88.085114,PhysRevLett.134.130403,PhysRevB.108.214429,PhysRevE.92.022115,BCS}. Similarly, various dualities for classical out-of-equilibrium many-body systems, described by continuous-time Markov processes, have been appreciated for decades—for example, the self-dualities of exclusion processes~\cite{liggett1985interacting,schutz2001exactly,giardina2009duality,1997JSP....86.1265S} and reverse duality  \cite{Schutz2023}; see also the recent systematic approaches using Lie algebras~\cite{giardina2026duality}, Hecke algebras and Macdonald polynomials \cite{CdGWduality}, and solvable vertex models \cite{Kuanduality}. The study of such equivalence relations not only provides mathematical elegance but also leads to substantial improvements in computational efficiency~\cite{lootens2025entanglement}.

Tensor networks provide a powerful framework for understanding non-trivial correlations in strongly interacting many-body systems~\cite{RevModPhys.93.045003}. For instance, using tools from quantum information theory, it has been shown that the ground states of large classes of one-dimensional local Hamiltonians, $H = \sum_{i} h_{i,i+1}$, are not structureless wavefunctions in the underlying Hilbert space. Rather, they reside in a corner of the exponentially large Hilbert space and are efficiently described by matrix product states (MPS) of the form $\sum_{ \{ \tau_n \}} \Tr\left( \mathbf{A}^{\tau_1}_1\cdots \mathbf{A}^{\tau_N}_N \right) | \tau_1 \dots \tau_N \rangle$, where $\mathbf{A}$ is a rank-three tensor. More specifically, this class enjoys the frustration-free property $h_{i,i+1} \left( \mathbf{A}_i \mathbf{A}_{i+1}\right)= 0$, which ensures that the true many-body ground state is annihilated by every local term of the Hamiltonian. Exploiting the strong parallel between quantum mechanics and Markov processes~\cite{alcaraz1994reaction}, properties of MPS in quantum systems often generalise directly to their stochastic counterparts, where the closest analogy for the frustration-free condition is detailed balance. Detailed balance is typically violated in one-dimensional boundary-driven local Markov processes, giving rise to richer phenomena than in their equilibrium counterparts~\cite{liggett1985interacting}; the steady states are not described by the Gibbs–Boltzmann measure and support non-vanishing currents. 

The seminal work of Derrida, Evans, Hakim, and Pasquier (DEHP)~\cite{derrida1993exact,uchiyama2004asymmetric,blythe2007nonequilibrium, PhysRevLett.104.210502} identified a class of such Markov processes in which detailed balance is strongly broken, yet the exact steady states can still be constructed using a matrix product ansatz (MPA). These steady states live in another corner of Hilbert space compared to the ground states of frustration-free models. This construction was achieved by imposing the bulk cancellation mechanism
$h_{i,i+1} \left( \mathbf{A}_i \mathbf{A}_{i+1}\right)
= \mathbf{A}_i \bar{\mathbf{A}}_{i+1}-\bar{\mathbf{A}}_i\mathbf{A}_{i+1}$,
together with additional appropriate boundary algebra, where $\bar{\mathbf{A}}$ is another rank-three tensor. Each local term of the Markov generator thereby induces an algebraic relation when acting on the steady state, and summing over sites produces an exact cancellation.

Similarly, matrix product descriptions of operators have been widely employed in strongly correlated systems, e.g., in the study of symmetries and duality transformations that intertwine the Hamiltonians of two dual models. Duality operators are ubiquitous, especially in systems with generalised symmetries. Matrix product operators (MPO) provide a systematic framework for dualities of quantum spin chains with generalised symmetries~\cite{PRXQuantum.4.020357, aasen2016topological}, encompassing many translationally invariant Markov processes~\cite{mussawisade1998branching,sfairopoulos2025multicriticality}; within this framework, dualities are implemented \textit{locally}—each local term of the Hamiltonian $h_{i,i+1}$ is mapped to its dual counterpart. As a canonical example, the Kramers–Wannier duality relates the two transverse-field Ising Hamiltonians $\sum_{i}(-J \sigma^{z}_{i} \sigma^{z}_{i+1} - g \sigma^x_{i})$ and $\sum_{i}(- J\sigma^{x}_{i} - g \sigma^{z}_{i} \sigma^{z}_{i+1})$, where $\sigma^{k}$ denote spin-$1/2$ Pauli matrices, $g$ and $J$ are parameters. The corresponding duality operator admits a MPO representation, $\otimes_{i} (\text{H}_{i} \text{CZ}_{i,i+1})$ (up to a half-site translation), where $\text{H}$ is the Hadamard gate and $\text{CZ}$ the controlled-Z gate. Each local term $\sigma^{z}_{i} \sigma^{z}_{i+1}$ ($\sigma^{x}_{i}$) is mapped to local term of the dual model $\sigma^{x}_{i}$ ($\sigma^{z}_{i} \sigma^{z}_{i+1}$).

In this Letter, we lift the MPA to higher rank and establish a novel MPO framework for duality operators of Markov processes that do not satisfy detailed balance and need not possess generalised symmetries. Dualities are implemented \textit{globally}, such that the conventional local duality relations are violated~\cite{PRXQuantum.4.020357,aasen2016topological}, but dualities of the whole system emerge through a cancellation mechanism of MPO. Our construction bridges spectral properties and physical observables. It is particularly powerful when one of the dual processes is in equilibrium, as calculations performed using the Gibbs–Boltzmann measure yield direct information about the out-of-equilibrium system~\cite{RevModPhys.87.593,PhysRevLett.99.150602,frassek2020duality}.

\emph{\textbf{Out-of-equilibrium MPO intertwiners.}}—\begin{figure}[t!]
    \centering
    \includegraphics[width=1\columnwidth]{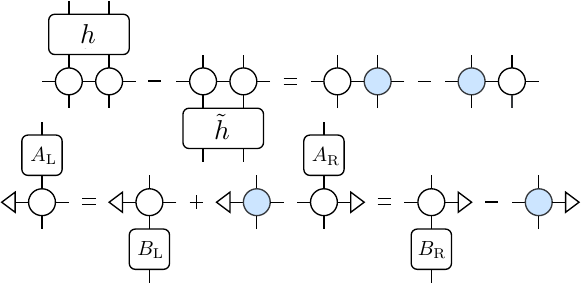}
    \caption{Diagrammatic illustration of the out-of-equilibrium MPO intertwiner,
    see Eqs. (\ref{eq:MPO-bulk}) and (\ref{eq:MPO-boundaries}) for the details.}
    \label{fig:fig1}
\end{figure}We consider a one-dimensional lattice consisting of $N$ sites, where each site $\tau_i$ can host a finite number $d$ of states. The Hilbert space is the $N$-fold tensor-product vector space $\mathcal{H}$ of total dimension $d^N$. Let $H^{T}$ be the generator of a Markov process and $H$ be the stochastic Hamiltonian. The time evolution of the many-body probability distribution is governed by the Master equation
$d | P(\tau_1 \dots \tau_N) \rangle /dt = H |  P(\tau_1 \dots \tau_N) \rangle$. Given two processes defined by 
\begin{equation}
H_1=A_\text{L}+\sum_{i=1}^{N-1} h_{i,i+1}+A_\text{R}
\label{eq:process1}
\end{equation}
and
\begin{equation}
H_2=B_\text{L}+\sum_{i=1}^{N-1} \tilde{h}_{i,i+1}+B_\text{R},
\label{eq:process2}
\end{equation}
where the bulk terms $h_{i,i+1}$ and $\tilde{h}_{i,i+1}$ acting only on adjacent sites. In addition, there are stochastic boundary terms acting on the first and last sites, denoted by the subscripts $L$ and $R$. We say that $H_1$ and $H_2$ are dual processes if there exists the following intertwining relation  
\begin{equation}
H_1 \mathcal{G} = \mathcal{G} H_2.
\label{eq:duality}
\end{equation}
The duality operator $\mathcal{G}$ relates two exponentially large operator spaces. We construct an efficient representation of the duality operator in the standard MPO form,
\begin{equation}
\mathcal{G} = \sum^{d}_{{ \substack{\tau_1, \dots, \tau_N = 1 \\ \tau'_1, \dots, \tau'_N = 1}}}
\langle W | \mathbf{L}^{\tau_1 \tau'_1}_1\cdots \mathbf{L}^{\tau_N \tau'_N}_N | V \rangle | \tau_1 \dots \tau_N \rangle \langle \tau'_1 \dots \tau'_N| .
\label{eq:MPO-form}
\end{equation}
Here, $\mathbf{L}$ is a rank-four tensor, and $W$ ($V$) denotes the left (right) boundary vector. Furthermore, we impose that the MPO intertwiner satisfies the following algebraic relations in the bulk, which we refer to as the out-of-equilibrium generalised exchange relation (or generalised pulling-through relation using tensor-network terminology),
\begin{equation}
h_{i,i+1} \mathbf{L}_i\mathbf{L}_{i+1} - \mathbf{L}_i\mathbf{L}_{i+1} \tilde{h}_{i,i+1} =\mathbf{L}_i \mathbf{Z}_{i+1}-\mathbf{Z}_i\mathbf{L}_{i+1},
\label{eq:MPO-bulk}
\end{equation}
where $\mathbf{Z}$ is another rank-four tensor. 
In addition, two boundary algebraic relations are required,
\begin{equation}
\begin{split}
& \langle W | \left( A_\text{L} \mathbf{L} - \mathbf{Z} \right) =\langle W|\mathbf{L} B_\text{L}, \\
& \left( A_\text{R} \mathbf{L} + \mathbf{Z} \right)|V\rangle= \mathbf{L} |V\rangle B_\text{R},
\end{split}
\label{eq:MPO-boundaries}
\end{equation}
see Fig.~\ref{fig:fig1} for a diagrammatic representation. This novel set of generalised exchange relations consists of the bulk part as well as non-trivial boundary parts. Unlike in conventional local duality, the bulk term $h_{i,i+1}$ is not mapped to the dual bulk term $\tilde{h}_{i,i+1}$ when pulled through the tensor $\mathbf{L}_i\mathbf{L}_{i+1}$; instead, it generates a local divergence of tensors. The boundary relations are chosen such that the divergence is cancelled and the boundaries of the two processes $A_{L,R}$ and $B_{L,R}$ are interchanged.

We note that the bulk relation takes the form of the Sutherland equation \cite{sutherland1970two} for $h_{i,i+1} = \tilde{h}_{i,i+1}$, which arises from the Yang-Baxter equation, cf. \cite{Crampe:2014aoa}. In analogy, the boundary relations can be seen as a consequence of the boundary Yang-Baxter equation or equivalently as the lifting of the Ghoshal-Zamolodchikov relations to a higher rank \cite{Ghoshal:1993tm,SaWa}, see also \cite{Cantini2015,Cantini2016}.

We now show that these generalised exchange relations indeed imply the duality transformation Eq.~(\ref{eq:duality}). We first consider pulling all bulk terms through the MPO, $( \sum_{i=1}^{N-1} h_{i,i+1} ) \mathcal{G}$. Using the bulk relation, this produces
$
\sum_{i=1}^{N-1}\langle W|\mathbf{L}_1\cdots \mathbf{L}_{i-1}(\mathbf{L}_{i}\mathbf{Z}_{i+1} - \mathbf{Z}_i \mathbf{L}_{i+1})\mathbf{L}_{i+2}\cdots \mathbf{L}_N|V\rangle + \mathcal{G} ( \sum_{i=1}^{N-1} \tilde{h}_{i,i+1} )$,
which forms a telescoping sum of operators, leaving two residual boundary contributions. Next, we pull through the remaining boundary terms and use the boundary relations to generate additional terms, which cancel precisely with the remaining terms
$A_\text{L} \mathcal{G} + A_\text{R} \mathcal{G} = \langle W|\mathbf{Z}_1 \mathbf{L}_2\cdots \mathbf{L}_N|V\rangle - \langle W|\mathbf{L}_1\cdots \mathbf{L}_{N-1} \mathbf{Z}_N|V\rangle + \mathcal{G} B_{\text{L}} +\mathcal{G} B_{\text{R}}$.
We thus establish the global duality via the out-of-equilibrium pulling-through mechanism.

\emph{\textbf{Symmetric Simple Exclusion Process.}}—\begin{figure}[t!]
    \centering
    \includegraphics[width=1\columnwidth]{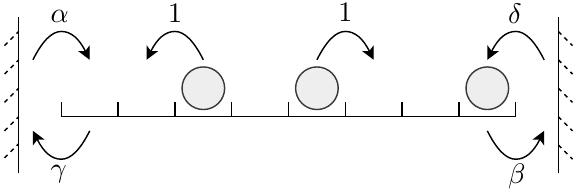}
    \caption{Illustration of the SSEP. Arrows (grey circle) indicate the allowed stochastic rules (particle).}
    \label{fig:fig2}
\end{figure}
In the following, we demonstrate the out-of-equilibrium MPO intertwiner for the symmetric simple exclusion process (SSEP), a canonical model for studying non-equilibrium transport that is known to be related to the integrable XXX Heisenberg chain \cite{schutz2001exactly,Crampe:2014aoa} 
The model describes stochastic hopping of hard-core particles. Each site of the one-dimensional lattice is either empty ($\tau_i = 1$) or occupied by a hard-core particle ($\tau_i = 2$), corresponding to $d=2$. Each particle hops to either adjacent site with probability $dt$, provided the target site is empty. In addition, the lattice is coupled to two particle reservoirs. A particle is added to the first (last) site with probability $\alpha dt$ ($\delta dt$) if the site is empty, and removed from the first (last) site with probability $\gamma dt$ ($\beta dt$) if it is occupied. We denote the four stochastic boundary parameters by $\{ \alpha_i \}_{i=1,\dots,4} =\{\alpha,\beta,\gamma,\delta\}.$ Out-of-equilibrium behaviour arises for generic boundaries satisfying $\alpha \beta - \gamma \delta \neq 0$. In the tensor-product basis of $\tau_i$, the out-of-equilibrium stochastic Hamiltonian $H_{\text{NE}}$ is given by
\begin{equation}
    h_{i,i+1} = \begin{pmatrix}
0 & 0 & 0 & 0 \\
0 & -1 & 1 & 0\\
0 & 1 & -1 & 0\\
0 & 0 & 0 & 0
\end{pmatrix},
\end{equation}
and the boundary terms are
\begin{equation}
A_{\text{L}} = \begin{pmatrix}
-\alpha & \gamma  \\
\alpha & -\gamma \\
\end{pmatrix} \quad \quad A_{\text{R}} = \begin{pmatrix}
-\delta & \beta  \\
\delta & -\beta \\
\end{pmatrix}.
\end{equation}

 We first consider the MPO realisation of mapping from the out-of-equilibrium process $H_1= H_{\text{NE}}$ to an equilibrium process $H_2 = H_{\text{E}}$ via Eq. (\ref{eq:duality}). $H_{\text{E}}$ will be determined explicitly below. This map coincides with the one obtained in \cite{frassek2020duality} using a perturbative approach \cite{alcaraz1994reaction}. The duality $\mathcal{G} = G $ defined by Eq. (\ref{eq:MPO-form}) and the generalised exchange relations Eqs.~(\ref{eq:MPO-bulk}) and (\ref{eq:MPO-boundaries}) are realised by
\begin{equation}
  \mathbf{L}  = L Y^{-1}
\quad 
    L = \begin{pmatrix}
-F & E \\
F & D \\ 
\end{pmatrix}
\quad \mathbf{Z} = \begin{pmatrix}
    0 & -\mathbb{1}\\
    0 & \mathbb{1}\\
\end{pmatrix}
\end{equation}
and $h_{i,i+1} = \tilde{h}_{i,i+1}$, where $Y$ is a product operator defined below, and $F$, $E$, and $D$ satisfy the following bulk algebra
\begin{equation}
[E,F] = F \quad [D,F] = -F \quad [D,E] = D+E.
\label{eq:bulk-algebra}
\end{equation}
In addition, the boundary algebra reads $\langle W | (\alpha E - \gamma D) = \langle W |$ and $(\beta D - \delta E )| V \rangle = | V \rangle$. We find the following bidiagonal \textit{bi-infinite-dimensional} matrix representation of this algebra, the non-vanishing matrix elements are 
\begin{equation}
\begin{split}
    &D_{n,n} = -E_{n,n} =  n  + \frac{1}{\beta + \delta} \quad F_{n,n+1} = \frac{1}{r_{n}} \\
    &D_{n+1,n} = \frac{\delta(\alpha+\gamma)r_n}{\delta+\beta}
    \quad E_{n+1,n} = \frac{\beta(\alpha+\gamma)r_n}{\delta+\beta}\\
    &r_{n} = \left( \frac{\beta + \delta}{\alpha \beta - \gamma\delta} \right) \left(\frac{1}{\alpha +\gamma} + \frac{1}{\beta + \delta} + n \right),
\end{split}
\label{eq:EandDmatrices}
\end{equation}
where ${n,m\in\mathbb{Z}}$. Using the standard orthonormal basis $\{|n\rangle\}_{n\in\mathbb{Z}}$ and $\langle n|m\rangle=\delta_{nm}$, the boundary vectors are given by $|V\rangle= |0 \rangle$ and $| W \rangle= \sum_{n\in\mathbb{Z}} |n \rangle$.

We emphasise that the special structure of this representation implies that the MPO can be realised exactly for \textit{finite} systems with matrix dimensions (bond dimensions) that are \textit{linear} in the system size and effectively given by $2N+1$. Matrix representations of the algebra are not unique; for instance, one may conjugate $\mathbf{L}$ and the boundary vectors by an arbitrary similarity transformation. Another \textit{infinite} matrix representation can be constructed~\cite{note1} by exploiting properties of Laguerre polynomials, reflecting the connections between MPS constructions and orthogonal polynomials~\cite{sasamoto1999one,uchiyama2004asymmetric}. A crucial difference from the bi-infinite matrix representations in Eq.~(\ref{eq:EandDmatrices}) is that $n$ is restricted to non-negative integers in~\cite{note1}.

Finally, we define the product operator $Y$ to complete the explicit MPO realisation of $G$. There are two equivalent constructions
\begin{equation}
Y_{\text{R}} = \begin{pmatrix}
-\frac{1}{\alpha + \gamma} & \frac{\beta}{\beta + \delta}\\
\frac{1}{\alpha + \gamma} & \frac{\delta}{\beta + \delta}\\
\end{pmatrix}^{\otimes N}
\quad
Y_{\text{L}} = \begin{pmatrix}
-\frac{1}{\beta + \delta} & \frac{\gamma}{\alpha + \gamma}\\
\frac{1}{\beta + \delta} & \frac{\alpha}{\alpha + \gamma}\\
\end{pmatrix}^{\otimes N}.
\label{eq:Ymatrix}
\end{equation}
For $Y = Y_{\text{R}}$, the dual $H_{\text{E}}$ is defined by
$B_{R} = A_{R}$ and $B_{L} = r A_{R}$, where $r= \frac{\alpha + \gamma}{\beta + \delta}$. For the alternative choice $Y_{\text{L}}$, the stochastic boundaries of $H_{\text{E}}$ become $B_{\text{L}} = A_{\text{L}}$ and $B_{\text{R}} = \frac{1}{r}A_{\text{R}}$. Notably, distinct $H_{\text{NE}}$ may be dual to the same $H_{\text{E}}$, leading to the closed MPO algebra discussed below.

Next, we establish the duality transformation from equilibrium to out-of-equilibrium. Consider another set of rates $\{ \alpha'_i \}_{i=1,\ldots,4}=\{\alpha',\beta',\gamma',\delta'\}$, and let $H_1 = H'_{\text{E}}$, $H_2 = H'_{\text{NE}} $ and $\mathcal{G} = G'$. The MPO intertwiner Eqs. (\ref{eq:MPO-form}) and generalised exchange relations Eqs.~(\ref{eq:MPO-bulk}) and (\ref{eq:MPO-boundaries}) are realised by
\begin{equation}
  \mathbf{L}  = Y \tilde{L}
\quad 
    \tilde{L} = \begin{pmatrix}
D' & -E' \\
F' & F' \\ 
\end{pmatrix}
\\
\quad  \mathbf{Z} = -\begin{pmatrix}
    \mathbb{1} & \mathbb{1}\\
    0 & 0\\
\end{pmatrix}. 
\end{equation} 
The bulk algebra remains unchanged Eq. (\ref{eq:bulk-algebra}), but the boundary relations turn into $\langle W' | (\alpha E' - \gamma D') = -\langle W' |$ and $(\beta D' - \delta E' )| V' \rangle = -| V' \rangle$, where $E'$, $D'$, and $F'$ are defined by Eq.~(\ref{eq:EandDmatrices}) evaluated at $\{ - \alpha'_i \}$, i.e., $\tilde{L}(\{ - \alpha'_i \})$. And $Y(\{ \alpha'_i \})$ is given by Eq. (\ref{eq:Ymatrix}). Note that $G'$ has the same computational complexity as $G$. Moreover, $G'$ reduces to the inverse of $G$ in the special case of $\{ \alpha_i = \alpha'_i \} $, and surprisingly the MPO structure is preserved under operator inversion.

Lastly, we complete the MPO intertwiner between two distinct out-of-equilibrium boundaries $\{ \alpha_i \}$ and $\{ \alpha'_i \}$ by composing $G$ and $G'$ associated with the equivalent $H_{\text{E}}$. Under the conditions $\alpha +\gamma= \alpha' + \gamma'$ and $\beta  + \delta  = \beta' + \delta'$, we define $\tilde{G}(\{\alpha_i\},\{\alpha'_i\}) = G(\{\alpha_i\}) Y(\{\alpha_i\}) Y^{-1}(\{\alpha'_i\}) G'(\{\alpha'_i\})$, which obeys $H_{\text{NE}} \tilde{G} = \tilde{G} H'_{\text{NE}}$. We remark that $\tilde{G}$ exhibits an important property from the theoretical perspective of tensor networks. The fusion of two MPOs generates a \textit{closed MPO algebra}
\begin{equation}
    \tilde{G}(\{\alpha_i\},\{\alpha'_i\}) \tilde{G}(\{\alpha'_i\},\{\alpha''_i\}) = \tilde{G}(\{\alpha_i\},\{\alpha''_i\})
\end{equation}
with the appropriate continuous boundary parameters. Remarkably, the bond dimension of this MPO algebra scales with system size, in contrast to conventional MPO algebras with a fixed bond dimension~\cite{RevModPhys.93.045003}.

\emph{\textbf{Physical Applications.}}—In the following, we discuss the physical applications of duality operators for the SSEP. We begin with the duality transformation of eigenstates and take $Y = Y_{\text{R}}$ WLOG. From the intertwining relation Eq.~(\ref{eq:duality}), one immediately obtains the mapping between eigenstates, $|P_1 \rangle = G |P_2 \rangle$, for the entire spectrum. In particular, the most physically relevant eigenstate is the steady state, defined by $H |P^{\text{ss}} \rangle  = 0$, which encodes the long-time behaviour and all static observables. For out-of-equilibrium boundaries, the exact steady state of $H_1$ is a highly correlated probability distribution supporting a non-vanishing current induced by the boundary drive. Its MPS representation reads
$|P^{\text{ss}}_1 \rangle = \sum^{2}_{\substack{\tau_1, \dots, \tau_N = 1}}  \langle W | \mathbf{A}^{\tau_1}_1\cdots \mathbf{A}^{\tau_N}_N | V \rangle | \tau_1 \dots \tau_N \rangle
$,
where $\mathbf{A}^{1} = E$ and $\mathbf{A}^{2}=D$. By contrast, for equilibrium boundaries, the steady state
$|P^{\text{ss}}_2 \rangle  = \frac{1}{\beta + \delta} \begin{pmatrix}
\beta  \\
\delta \\
\end{pmatrix}^{\otimes N}$
is given by a Bernoulli measure and therefore exhibits no spatial correlations. It is straightforward to verify that $|P^{\text{ss}}_1 \rangle = G |P^{\text{ss}}_2 \rangle$.

We next consider the evaluation of physical observables in the steady state. In the MPS framework, local observables are extracted from the transfer matrix constructed from $\mathbf{A}$, which requires spectral information of an infinite-dimensional matrix in the out-of-equilibrium case. By contrast, for the equilibrium process, the product structure of the steady state allows direct contraction with observables. By further exploiting the MPO intertwiner, one can access out-of-equilibrium properties via computations performed in the Gibbs–Boltzmann ensemble. As a concrete example, consider the multi-point density correlation function in the out-of-equilibrium steady state, $\langle + | n_i \cdots n_j |P^{\text{ss}}_2 \rangle$, where $n_i = \frac{1}{2}(\mathbf{1} - \sigma^{z}_i)$ and $\langle + |$ denotes the uniform probability vector such that $\langle + | H = \langle + |$. This quantity can equivalently be computed by averaging over the uncorrelated Bernoulli measure,
$\langle + | \tilde{n}_{i\cdots j} |P^{\text{ss}}_1 \rangle$,
where
$\tilde{n}_{i\cdots j}  = \langle W | \mathbf{L}_1\cdots \mathbf{X}_i \cdots\mathbf{X}_j \cdots \mathbf{L}_N | V \rangle$
and 
$\mathbf{X} = \begin{pmatrix}
    0 & 0 \\
    F & D \\
\end{pmatrix}$,
thereby recovering the standard result~\cite{uchiyama2004asymmetric,frassek2020eigenstates}.

\emph{\textbf{Discussion.}}—We have established a matrix product approach to the duality operator for boundary-driven Markov processes. By introducing a novel set of generalised exchange relations, we lift the DEHP Ansatz to operators and apply the telescopic cancellation mechanism to implement the duality transformations. Out-of-equilibrium dualities are realised globally, in contrast to local dualities associated with generalised symmetries. The MPO intertwiner establishes a spectral equivalence between distinct Markov processes with stochastic boundaries and provides a direct correspondence between their eigenstates and observables, including the surprising mappings between out-of-equilibrium and equilibrium. A related duality was previously explored in the hydrodynamic approximations of SSEP and described as a ``miracle'' in Ref~\cite{PhysRevLett.99.150602}, and understood using changes of (field) variables~\cite{tailleur2008mapping}; our approach provides an exact many-body lattice realisation of this duality using MPO. The intimate connection to quantum integrability is explored in a separate companion work~\cite{tbp2026}, where an MPO is derived from integrable defects   \cite{Bajnok:2006bf}. We hope to report analogous constructions for the asymmetric simple exclusion process and XXZ in future work~\cite{schutz2001exactly, PhysRevLett.95.240601,Rafael,gehrmann2025exact}. Another direction is to investigate the application of our construction to open quantum systems~\cite{prosen2015matrix} as well as other perspectives on the exchange relations~\cite {fendley2025xyz,rubio2026local}.

\emph{\textbf{Acknowledgements.}}—We thank Cristian Giardinà for helpful discussions. JL acknowledges discussions on related topics with
Paul Fendley, Juan P. Garrahan, Robert Jack, Hosho Katsura, Katja Klobas, Linhao Li, Laurens Lootens, and Wei Tang. RF thanks Zolt\'{a}n Bajnok and Istv\'{a}n M. Sz\'{e}cs\'{e}nyi for very inspiring discussions on isospectral spin chains, defects and matrix product operators during an early stage of this project. JL and FV acknowledge support by the UKRI Grant No. EP/Z003342/1. JL and FV thank the Isaac Newton Institute for Mathematical Sciences, for support and hospitality during the programme ``Quantum Field Theory with Boundaries, Impurities, and Defects" where work on this paper was undertaken. This work was supported by EPSRC grant No. EP/R014604/1. JdG was supported by the Australian Research Council. RF was supported by INdAM (GNFM),
the FAR UNIMORE project CUP-E93C23002040005, and by the PRIN project CUP-E53D23002220006.

\bibliography{SSEPrefs}
\end{document}